\documentstyle[12pt,psfig]{l-aa}
\def\P{Paczy\'{n}ski~}
\newcommand{\kms}{\mbox {km s$^{-1}$}}
\def\apgt{\ {\raise-.5ex\hbox{$\buildrel>\over\sim$}}\ }
\def\aplt{\ {\raise-.5ex\hbox{$\buildrel<\over\sim$}}\ }
\newcommand{\mbh}{\mbox {${M}_{\rm BH}$}}
\newcommand{\pyr}{\mbox {{\rm yr$^{-1}$}}}
\newcommand{\rs}{\mbox {$R_{\odot}$}}
\newcommand{\ms}{\mbox {$M_{\odot}$}}
\newcommand{\ace}{\mbox {$\alpha_{ce}$}}
\newcommand{\md}{\mbox {$\dot{M}$}}

\newcommand{\myr}{\mbox {~${\rm M_{\odot}~yr^{-1}}$}}
\newcommand{\es}{\mbox {~erg s$^{-1}$}}
\def\etal{{et al.}\ }

\def\eg{{\it e.g.}\ }
\def\apj{ApJ}

\def\pppartf#1#2#3{{{\partial^2{#1}}\over{\partial 
{#2}}{\partial {#3}}}}

\begin{document} 

\thesaurus{08.02.1, 02.01.2, 08.23.2} 

\title{Cyg X-3: can the compact object be a black hole?} 
\author{Ene Ergma \inst{1,3}  
\and
Lev R. Yungelson \inst{2,3}}
\institute{Physics Department, Tartu University, \"Ulikooli 18, EE2400
Tartu, Estonia (ene@physic.ut.ee)
\and
Institute of Astronomy of the Russian Academy of Sciences, 48
Pyatnitskaya Str., 109017 Moscow, Russia (lry@inasan.rssi.ru)
\and 
Astronomical Institute ``Anton Pannekoek'', University of Amsterdam,
Kruislaan 403, 1098 SJ Amsterdam, The Netherlands} \offprints{E.Ergma}

\date{Received;accepted}

\maketitle \markboth{E. Ergma \& L. Yungelson: \, Cyg X-3}{} 
\begin{abstract}
By means of population synthesis we find that the expected Galactic number
of black holes with massive helium star companions is $\sim 100$ and
depends on an assumed threshold for $M_{\rm pre-BH}$.  ~The overwhelming
majority of these systems has orbital periods in excess of 10\,hr, with a
maximum at $\sim 100$\,hr, while under the Illarionov \& Sunyaev (1975)~
disk formation criteria for accretion from the strong stellar wind of
Wolf-Rayet star disk accretion is possible only for orbital periods below
$\sim 10$\,hr. However, the number of such short-period systems is
vanishingly small. If the accretor in Cyg X-3 is a 10\,\ms\ black hole,
then the accretion rate will be super-Eddington. Super-Eddington accretion
may be responsible for the formation of jets in Cyg X-3 and may also
support an X-ray luminosity as high as $\sim 10^{39}$ \es.  From the
orbital period distribution for neutron stars with massive helium
companions we find that if during the common envelope phase a neutron star
accretes at $\md_{\rm Edd}~$ and spins-up to the equilibrium period, then
in most systems it acts as a``propeller'' and accretion from the WR star
wind is impossible.  For the model with two massive helium stars as an
immediate progenitor of Cyg X-3, requirement of accomodation of two WR
stars in the post-common-envelope orbit combined with severe mass loss by
them prevents formation of BH+WR systems with orbital periods less than
several days. 
\end{abstract}


\section{Introduction}

At present, three black hole candidates with massive early-type
companions are known (Cyg X-1, LMC X-3 and LMC X-1).  About half a
dozen of black hole candidates with low-mass companion stars have been
identified as the so-called X-ray transients (Tanaka \& Lewin 1995; Tanaka \&
Shibazaki 1996). Yet another candidate may be
the puzzling 4.8\,hr 
orbital period X-ray binary Cyg X-3~ 
(Giacconi \etal 1967). Its spectrum is hard, with a 
tail extending to ultra-high energy $\gamma$-rays (Lloyd-Evans \etal
1983). Cyg X-3 
shows an asymmetric sinusoidal modulation in the low-energy X-ray
emission (Mason \etal 1976). The same modulation is present in 
the infrared light curve
(Becklin \etal 1973; Mason \etal 1986). Infrared spectroscopy
has recently shown that the companion is a Wolf-Rayet star of the WN7
subtype (van Kerkwijk \etal 1992, 1996; van Kerkwijk  1993). Cyg X-3
has huge radio outbursts with evidence for jetlike emission expanding at
$\sim 1/3$ of the light velocity (Gregory \etal 1972; Geldzahler \etal 
1983). 

The orbital period of Cyg X-3 is increasing on a time scale of
850\,000 years (Kitamoto \etal 1995). With this $\dot P / P$~ it
is possible to estimate the stellar wind mass loss rate as $\dot {M}_{\rm
dyn} = M_{\rm tot} (\dot {P}/ 2 P) \approx 6 \times 10^{-6}
(M_{\rm tot}/10 \ms)$\,\myr, where $M_{\rm tot} = M_{\rm c} +
M_{\rm WR}$, $M_{\rm c}$~ and $M_{\rm WR}$~ are the masses of
the compact object and the WR star, respectively.

Another estimate of the mass loss rate can be made using the
infrared flux $F_{\rm ff}$. Van Kerkwijk \etal (1996) estimated
$\dot {M}_{\rm ff}$~ using a simple relation $\dot {M}_{\rm ff}
/ v_{\rm w} \propto F_{\rm ff}^{3/4}$. For the
wind velocity of $1450 \pm 150$ \kms\  and $F_{\rm ff}$
determined from  the May 1992 infrared light curve
they find $\dot {M}_{\rm ff} \approx  1.2 \times 10^{-4}$ \myr,
an order of magnitude larger than the dynamical estimate
$\dot{M}_{\rm dyn}$~ (for $M_{\rm WR} = 10$ \ms, $M_{\rm c} =
1.4 \ms$). But as was already discussed by van Kerkwijk \etal
(1996), $\dot {M}_{\rm ff} \sim 10^{-4}$\,\myr\ that is inferred
from the infrared flux should be regarded as highly uncertain.
Due to  clumpiness of the wind, the true mass loss rate may be
lower than $\dot {M}_{\rm ff}$ by a factor 2--3 and for Cyg
X-3 the difference may be even larger (see also Cherepashchuk \&
Moffat (1994) and references therein for a discussion of the
effect of clumps).

Cyg X-3 is a strong X-ray source with mean X-ray flux
$F_{\rm X}(2-20\, {\rm KeV}) \approx 6.8 \times 10^{-9} \es {\rm
cm^{-2}}$, i.e. $L_{\rm X} \approx 1.2 \times 10^{38}(d / 12 \
{\rm Kpc})^2$\,\es~ (Kitamoto \etal 1995).  Cherepashchuk \&
Moffat(1994) argue that since $L_{\rm bol}({\rm WR})
\sim 3 \times 10^{39}$ \es~ and because the effect of the X-rays is
observed in the IR range, the true intrinsic X-ray luminosity of
the accreting relativistic object in Cyg X-3 must be
considerably higher than the observed mean value of the hard
X-ray luminosity $L_{\rm X} \sim 10^{38}$ \es.
In their opinion, ``this fact favours the presence of an accreting black 
hole, as opposed to a neutron star, in Cyg X-3''. 

Recently,  Schmutz \etal (1996) observed time-variations in
the profiles of several infrared emission lines from Cyg X-3. They
concluded that the variations are due to the orbital motion of the WR
star and derive a mass function for Cyg X-3 of 2.3\,\ms. Assuming
reasonable values for the mass of WR star and the inclination angle, they
obtained a range of masses 7--40 \ms\ for the compact object, with the most
likely value of 17 \ms, from which they concluded that the compact
component of Cyg X-3 is a black hole. 

Van den Heuvel \& De Loore (1973) were the first to suggest that
Cyg X-3 may be a system in a post massive X-ray binary (MXRB)
stage and may harbour a compact object (neutron star or black
hole) and a helium star of several solar masses.  
Tutukov \& Yungelson (1973a) independently suggested that
the optical components of MXRB will in the later evolutionary stage become
WR stars provided that they are massive enough.

In the present paper we attempt  to model the Galactic population of 
black-hole + massive helium star systems and to show that the very short 
orbital period makes Cyg X-3 an extremely rare example of such systems 
detectable by X-ray observations. Also, we discuss the circumstances 
which prevent discovery of a large Galactic population of 
neutron star+helium star systems, which are probably quite numerous.

\section{Cyg X-3 - a unique system with a black hole accretor?}

\subsection{Formation scenarios for BH + WR binaries}

The calculations for the present study were made by means of the
population synthesis code which was previously applied by one of the
authors (LRY) for the modelling of different components of the binary star
population of the Galaxy, among them binary Wolf-Rayet stars (Yungelson \&
Tutukov 1991; Vrancken \etal 1991), neutron stars (Tutukov \& Yungelson
1993a,b) and high-mass X-ray sources (Iben \etal 1995). The basic
assumptions important for the present study are the following. 

The initial mass function of the primaries in binary stars is a 
power-law    $ dN \propto M_1^{-2.5}dM_1$.  
The initial distribution of binaries over separation $a$ is flat in $\log 
a$ for $0 \leq \log (a / \rs) \leq 6$. The orbits of close binaries are 
initially circular. 
The initial distribution of binaries over mass ratios of components $q_0 = 
M_2/M_1$ is flat for~ $0 < q_0 \leq 1$. Discussion of the effects  of the 
assumed distributions over $a$ and $q_0$ upon the population of high-mass 
binaries in the Galaxy may be found elsewhere (\eg\ Portegies Zwart \& 
Verbunt 1995; Portegies Zwart \& 
Yungelson 1997). They appear to be of no serious significance to the (mostly) 
qualitative conclusions of the present paper.

Normalization of the Galactic stellar birthrate function corresponds to
the formation of one binary system with the initial mass of the primary
greater than 0.8 \ms\ per year. The rate of binarity of stars is assumed
to be equal to 100 \%.\footnote{For another birthrate normalization and
assumptions on the rate of binarity all numbers of binaries given below
have to be rescaled.} ~With this normalization and the assumed rate of
binarity our model gives a combined rate of {\mbox SN Ib/c} and {\mbox SN
II}~ $\sim 0.02$ \pyr, consistent with the sum of expected rates of
Galactic {\mbox SN Ib/c} and {\mbox SN II}~(Cappellaro \etal 1997). 

Below, we use the following notation: MS - main-sequence star, RLOF -
stage of the Roche lobe overflow~(stable or with formation of a common
envelope), He - helium star remnant of a star which experienced a mass
loss (in fact, a WR\, star) , SN - supernova explosion, BH - black hole.
Then all basic scenarios for the formation of black-hole + helium star
binaries may be summarized as:  \begin{eqnarray} \lefteqn{{\rm A.~~~ MS_1
+ MS_2 \rightarrow RLOF_1 \rightarrow He_1 + MS_2 \rightarrow SN_1 + MS_2
}} \nonumber \\ \lefteqn{{\rm ~~~\rightarrow BH_1 + MS_2 \rightarrow
RLOF_2 \rightarrow BH_1 + He_2 }} \nonumber \end{eqnarray} \noindent and
\begin{eqnarray*} \lefteqn{{\rm B.~~~ MS_1 + MS_2 \rightarrow He_1 + MS_2
\rightarrow SN_1 + MS_2}} \nonumber \\ \lefteqn{{\rm ~~~\rightarrow BH_1 +
MS_2 \rightarrow BH_1 + He_2. }} \nonumber \end{eqnarray*} Scenario A
occurs in the systems in which stellar wind mass loss does not prevent
significant expansion of stars and RLOF.  In the opposite case scenario B
occurs. In fact, in scenario B components of binaries evolve independently
like single stars. In absence of hydrodynamic calculations of the mass
exchange, the upper mass limit of stars which may experience {\em
classical} RLOF is, in fact, one of the free parameters of the problem
(for discussion of this issue see, \eg\, Sybesma 1986; Moffat 1995).
Conventional computations of stellar evolution with stellar wind mass loss
rates suggested by observations (e.g. Massevich et al. 1979; Sybesma 1986;
Vanbeveren 1991) show that scenario A occurs in systems with the initial
mass of the primary component below 35 - 50 \ms.  However, depending on
the mass accumulated in the ${\rm RLOF_1}$~ event, the secondary in
scenario A may avoid RLOF and, {\em vice versa}, the secondary in scenario
B may encounter RLOF\, if its mass is sufficiently low. In the present
study we assume that the border between scenarios A and B is at $M_1 = 50$
\ms. Systems which evolve under the influence of stellar wind mass loss
have larger orbital separations in the ${\rm BH_1+He_2}$~ stage than
systems which evolved through nonconservative RLOF. Decrease of the
abovementioned border below $M_1 = 50$ \ms\ would make the conditions for
the formation of Cyg X-3-like systems even less favourable than is found
below (see Fig. 1). 

The mass of the helium star immediately 
after the end of mass exchange episode is given by the relation $M_{\rm He} = 
0.066 M_{\rm MS}^{1.54}$, which represents a fit to the evolutionary 
calculations by Tutukov \& Yungelson (1973b) and Iben \& Tutukov (1985). 

To estimate the variation in the separation of components in the 
common envelope stage we applied the equation suggested by Tutukov \& 
Yungelson (1979):
\begin{equation}
\frac {(M_{10}+M_2)(M_{10}-M_{1f})}{2a_0} \approx \ace M_{1f} M_2 \left(
\frac {1}{2a_f} - \frac{1}{2a_0}\right),
\end{equation}
where subscripts $0$ and $f$ refer to the initial and final states, 
respectively, and \ace\ is the parameter which measures the
ratio of the binding energy of the common envelope and the
change of the orbital energy of the binary (see e.g. Iben \&
Livio 1993; Livio 1996 for discussion).   
This equation implies that the spiral-in starts when the envelope
surrounding the two cores expands to $\sim 2 a_0$ and mass is actually
ejected from the gravitational potential of both components. 
Computations were performed for the common envelope parameter \ace = 1, 
which gave reasonable results in our previous attempts  to model 
the population of binary stars. When compared to the equation for $a_f /a_0$ 
derived by Webbink (1984), which is often applied for population synthesis 
studies, \ace = 1~ in Eq. (1) corresponds to $\ace \approx 4~$ in  
Webbink's equation. For the latter it formally means that most of the 
energy for 
expulsion of the common envelope comes from  sources other than the orbital 
energy of the binary. 
However, there are clear indications that with 
$\ace \sim 3-4$ (in Webbink's formulation) one would obtain more reasonable 
agreement with observations for several classes of 
descendants of massive binaries than with  lower values of \ace\ (Portegies 
Zwart \& Yungelson 1997).

\subsection{Masses of the Cyg X-3~ components}

The actual mass range of progenitors of black holes both in single stars
and in binaries is, in fact, unknown.  The main factors contributing to
this are the uncertain mass loss rates of hydrogen- and helium-rich stars,
poor understanding of mixing processes in stellar interiors and of
processes during a supernova explosion and a poorly known equation of
state (see \eg Woosley \etal 1995; Brown \etal 1996 and references
therein). The ``standard'' assumption is that binary components more
massive than 40\,\ms\ form black holes (van den Heuvel \& Habets 1984).
Recent calculations of Brown \etal seem to confirm this limit, while
according to Woosley \etal (1995), who assume an extremely high mass-loss
rate for helium stars, this limit may be as high as $\sim 60$\,\ms.
Another implication of the Woosley \etal results is convergence of
pre-supernovae masses to 3--4 \ms.  Actually, significant spread of masses
in observed candidate black holes in soft X-ray transients (4--16 \ms)
suggests that there may be factors other than the initial mass which
determine the fate of a star (Ergma \& van den Heuvel 1997). Under the
given circumstances, we treated the masses of black holes and their
progenitors as an additional model parameter. Computations were performed
on the assumption that black holes result from objects which have a mass
mass exceeding 7 or 10\,\ms\ at the end of the helium star stage. Under
the initial-final mass relation given in Sect. {\it 2.1}~ and using
Langer's (1989) mass loss rates for helium stars, these limits correspond
to $M_{ZAMS} \approx 30$ or 50\,\ms. We assumed that no mass is ejected
during a massive helium star remnant collapse into BH (Burrows 1987;
Woosley \& Weaver 1995).

The mass of the WR component of Cyg X-3~ is likewise uncertain. 
Cherepashchuk \& Moffat (1994) suggest the range 10--50 \ms\ based on the 
spread of estimates of masses of actually observed WN7 stars.
One can use
the estimated mass loss rate of the stellar wind to get an idea about the
WR star mass in Cyg X-3. From the observations for five binary Wolf-Rayet
stars, Abbott \etal (1986) derive
$\md_{\rm w} \propto M_{\rm WR}^{2.3}$. ~Langer (1989) suggests, from the 
fits of stellar models  to the observational data, that in the
WR star core helium burning phase $\dot {M}_{\rm w} \approx (1-6) \times
10^{-8}(M_{\rm WR}/ \ms)^{2.5}$ \myr. 
If  $\md_{\rm w} \sim 1/10 \md_{\rm ff}  \approx 1.2 \times 10^{-5}$ \myr\
(van Kerkwijk \etal 1996)
and $\dot {M}_{\rm w} \approx \dot {M}_{\rm dyn},$~ 
then $M_{\rm tot} \sim 20 \ms$. A direct application of Langer's 
formulae gives  $8 \aplt M_{\rm WR}/\ms \aplt 17$. 

The rate of accretion onto a black hole may be estimated as  
\begin{equation}
\dot{M}_{\rm acc} \approx \frac{\pi r_{\rm acc}^2}{4\pi
a^2}\dot{M}_{\rm w}, ~~~{\rm where}~~r_{\rm acc}=\frac{2G \mbh}{v_{\rm 
w}^2}.
\end{equation} 
Then, according to Kepler's third law  
\begin{equation}
\dot{M}_{\rm acc} \approx 0.14  \frac{M_{\rm BH}^2}{v_{1000}^4}
P_{\rm hr}^{\frac {4}{3}} (\mbh + M_{\rm WR})^{\frac{2}{3}}  
\dot{M}_{\rm dyn}, 
\end{equation} 
where all masses are in \ms, ~$P_{hr}$ is the orbital period in hours,  and 
$v_{1000}$~ is the wind velocity in 
1000~ \kms. For $ \mbh = 10 \ms $, $v_{1000} = 1.5$, $P_{\rm hr} = 4.8$,
$M_{\rm WR} = 10 \ms $
one  gets
$\md_{\rm acc} \approx 0.05\times \md_{\rm dyn} \approx 5.7 \times 10^{-7}$ 
\myr.\ The Eddington limit for the accreting black hole
(helium accretion and $R = 3 r_{\rm g}$) is~ $2.3 \times 10^{-8} (M_{\rm 
BH} / \ms)$ \myr\ and for $M_{\rm BH} = 10 \ms$ it is  $\md_{\rm Edd} 
\approx  2.3 \times 10^{-7}$ \myr.
Thus, in the latter case the accretion rate exceeds the Eddington limit. 
It is easy to find that if the accreting object is a neutron star, then 
\begin{equation}
\md_{\rm acc}({\rm NS})=\left( \frac{M_{\rm NS}}{M_{\rm BH}} \right)^2
\left( \frac{M_{\rm WR}+M_{\rm NS}}{M_{\rm WR}+M_{\rm BH}} 
\right)^\frac{2}{3} \dot{M}_{\rm acc}({\rm BH}).
\end{equation} 
For the same mass of the Wolf-Rayet star (10 \ms), but 
for a neutron star or low-mass black hole
 accretor (1.5 \ms, Brown 1995), it is simple to estimate by Eq. (4) that
$\dot{M}_{\rm acc}$ is below the Eddington rate.
However, if the accretion is super-Eddington, it may be
responsible for the formation of jets in Cyg X-3. A high accretion rate
may support an X-ray luminosity as high as $\sim 10^{39}$ \es\
which is consistent with Cherepashchuk \& Moffat's (1994) estimate. 

No kicks were imparted to the nascent BH. Even in the case of neutron
stars the problem of natal kicks is still unsolved:  observational data
may be explained both on the assumption of the absence of kicks (Iben \&
Tutukov 1996) and of their presence (van den Heuvel and van Paradijs
1997). As an argument for absence of kicks for BH one may consider the
fact that black hole candidates in low-mass X-ray binaries (LMXB) share
the radial velocities of their local rest frames in the Galaxy (with the
exception of Nova Sco 1994). The dispersion of their distances to the
Galactic plane is smaller than for LMXB with neutron stars by more than a
factor 2. This indicates that any kick velocities that the black hole
systems may have received at their formation are considerably smaller than
those of neutron star systems (White \& van Paradijs 1996).

\subsection{Model population of BH + WR and NS + WR binaries}

\begin{figure}[ht]
\psfig{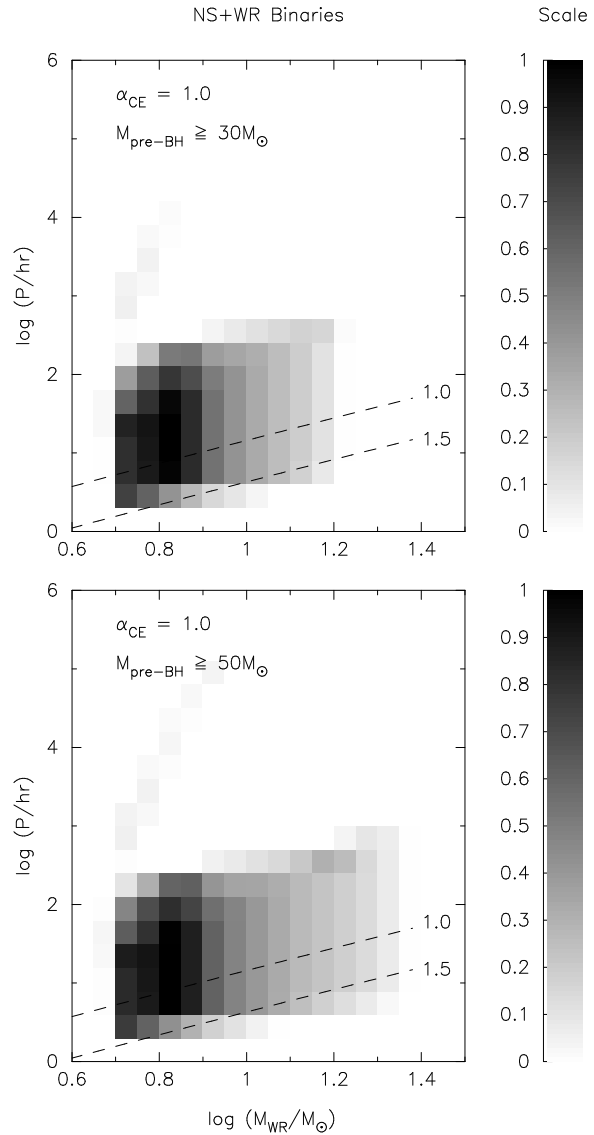}
\caption[]{Normalized number-density distribution of BH+WR systems
over orbital periods and masses of WR components 
for two assumptions on the minimum mass of the progenitors 
of black holes - 30 \ms\ (upper panel) and 50 \ms\ (lower panel). Dashed 
line labelled by 1.0 gives the upper limit of orbital periods for which 
disk accretion onto BH is possible if $M_{\rm BH} = 10\ms$ and $v_{\rm 
1000} = 1.0$. Maximum of the gray-scale is the same for both 
panels and corresponds to $\pppartf{N}{\log {\rm M}}{\log P} =
274,$~where $N$ is the number of systems. }
\end{figure}
\begin{figure*} 
\centerline{
\psfig{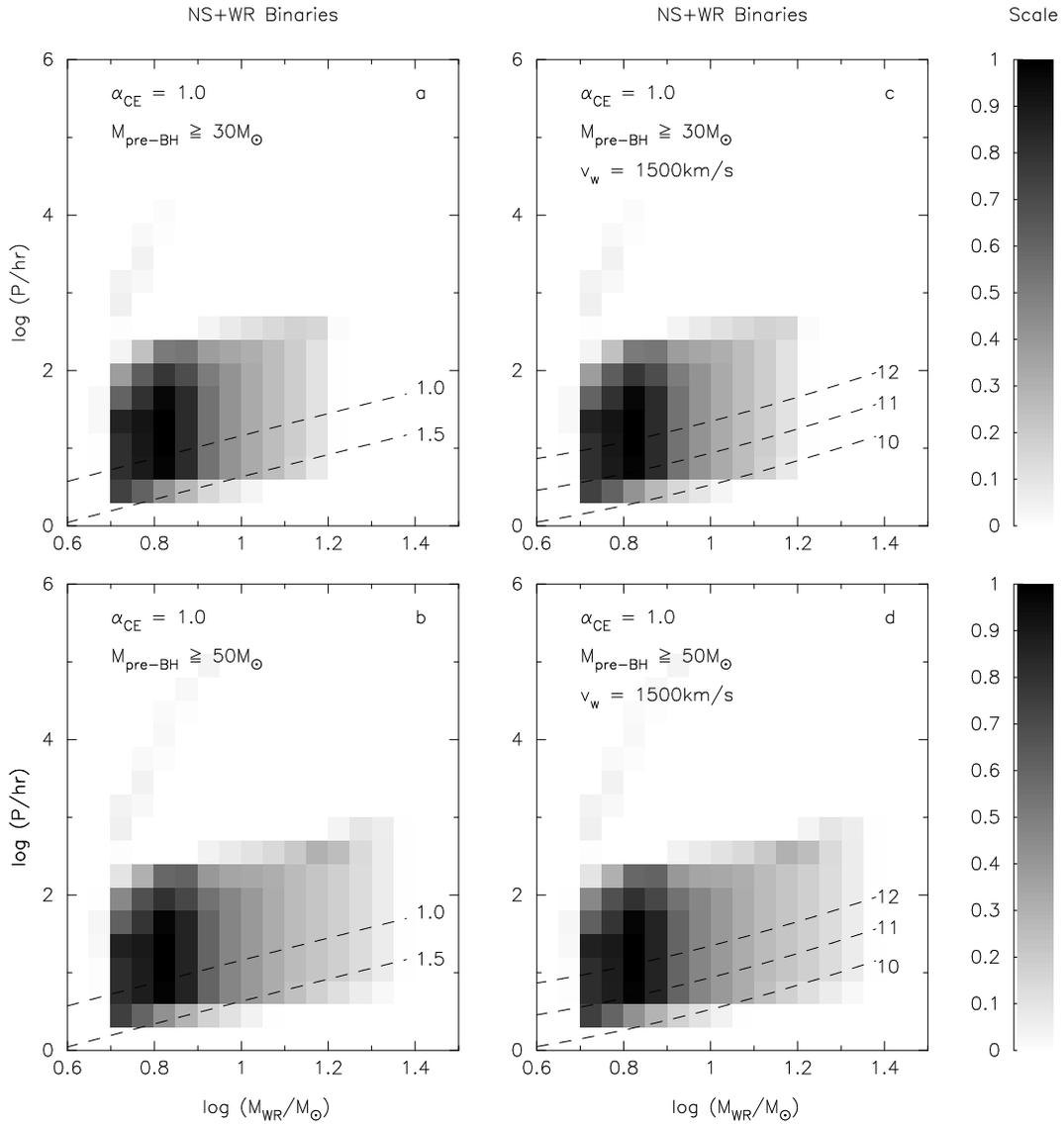} }
\caption[]{The same as in
Fig. 1, but for NS+WR systems. In panels (a) and (b) dashed lines labelled
by 1.0 and 1.5 give the upper limits of orbital periods for which
``propeller'' effect does not prevent disk accretion onto NS if $v_{\rm
1000} = 1.0~ {\rm and~} 1.5$~ respectively. In panels (c) and (d) dashed
lines labelled 10, 11, and 12 give the upper limits of the the orbital
periods for which spin deceleration timescale of a neutron star is shorter
than the helium-burning time of its companion if magnetic field strength
is $B = 10^{10},~ 10^{11}$, ~and $10^{12}$~ {\rm Gs}, respectively. 
Maximum of the gray-scale is the same for all panels and corresponds to
$\pppartf{N}{\log {\rm M}}{\log P} = 1160$. } 
\end{figure*}

Figure 1 shows the distribution of BH + WR binaries over orbital periods
and masses of WR stars for two assumptions on the progenitor mass of the
BH.  The minimum mass of the WR star is taken equal to 5\ms. The
difference between the two distributions may be due to the fact that for
$M_{\rm pre-BH} \geq 30$ \ms\ both scenarios A and B operate, while for
$M_{\rm pre-BH} \geq 50$ \ms\ only scenario B is effective, i.e. in the
first case some of pre-WR stars were able to accumulate some mass via
stable RLOF. The predicted numbers of BH + WR stars in the two cases are
$\sim 300$ and $\sim 100$, respectively. These numbers can be considered
only as giving an order of magnitude estimate on the potential number of
the Galactic BH+WR systems. If natal kicks or mass ejection accompany the
formation of BH, they may reduce the number of these systems, but they
cannot be overly significant for the most massive binaries we are dealing
with. 
 
More important, in our opinion, is the fact that the overwhelming majority
of these systems have orbital periods longer than 10 hours, with a maximum
of $\sim 100$ hours. This may be helpful in to understanding why Cyg X-3
is unique.  On the basis of the seminal Illarionov \& Sunyaev's (1975)
paper "Why the Number of Galactic X-Ray Stars Is So Small?"one can make
the following estimate.  Illarionov \& Sunyaev have shown that in the case
of accretion of stellar wind matter in a detached binary system the
specific angular momentum of the matter captured by the relativistic star
is typically small. Therefore, no accretion disk is formed around the
relativistic star. Consequently, very special conditions are required for
a black hole in a detached binary system to be a strong X-ray source. A
disk may form if the specific angular momentum of accreting matter exceeds
the specific angular momentum of the particle in the last inner stable
orbit with $ R = 3 r_{\rm g} $ which is equal to $Q_{\rm min} = \sqrt {3}
r_{\rm g} c$. Thus, the disk formation criterion is
\begin{equation}
\frac{1}{4}\Omega R_{\rm A}^2 > \sqrt {3} r_{\rm g} c,
\end{equation} 
where $R_{\rm A}$ is
the capture radius of  stellar wind matter by the relativistic star. This
condition is fulfilled when the period of binary system $P_{\rm orb} < 4.8
(M_{\rm BH} / \ms) v_{1000}^{-4} {\delta}^2$~ hr,  where $\delta 
\sim 1$ is a dimensionless parameter. Typical Wolf-Rayet stellar wind
velocities range from 1000 \kms\ to 4000 \kms\ (Conti 1988). As a 
result, disk accretion is possible only for systems
having a very short orbital period
(Fig.~1). Curiously enough, if $M_{\rm BH} \approx 5\ms$\ and  
wind velocity $\sim  1500$ \kms\ for Cyg X-3,  the disk accretion condition
is satisfied when
$P_{\rm orb} \approx 4.8$ hr! (A similar conclusion was reached by Iben 
\etal 1995).

A Wolf-Rayet binary with $P_{orb} = 4\d32$ and suspected BH-companion 
{\mbox HD~197406} (Cherepashchuk 1991; Marchenko \etal\ 1996)
may be a representative of the  
population of BH+WR binaries which do not show up as X-ray sources.  
   
The population of BH + WR stars can be compared with the population of 
neutron stars accompanied by WR stars (henceforth, NS + WR). Figure 2 
shows $\log M_{\rm WR} - \log P$ plots for NS + WR stars similar to Fig. 1.
The respective numbers of systems are $\sim 570$ and $\sim 700$ for 
$M_{\rm pre-NS} \leq 30~ {\rm and}~ 50$ \ms.
These numbers provide upper limits for this possible population because no 
natal kicks were imparted to   neutron stars. It is also
worthwhile to note that natal kicks tend to 
influence more strongly the eccentricities of orbits of post-supernovae 
systems than their orbital periods.

One may ask why NS + WR systems are not observed. We would like to call
attention to the following.  If the WR component is formed via RLOF one
can expect that a common envelope forms because of the high mass ratio of
components at the instant of the RLOF by the progenitor of the WR star. As
a rule, the common envelope stage lasts for $\sim 10^3 - 10^4$ yrs. During
this stage the NS may accrete at the Eddington's rate of $1.5 \times
10^{-8}$ \myr. Accretion spins up the neutron star (Pringle \& Rees 1972; 
Davidson \& Ostriker 1973) to the equilibrium period
\begin{equation}
P_{\rm rot} = P_{\rm eq} \approx 4 B_{11}^{\frac{6}{7}} 
\md_{11}^{-\frac{3}{7}} M_{\rm NS}^{-\frac{5}{7}}~~~{\rm s},
\end{equation}
where $B_{11} = B/10^{11}$ G, $\md_{11} = \md / 1.5 \times 10^{-11}$ 
\myr, $M_{{\rm NS}}$ is the neutron star mass in \ms. The amount of matter 
required for the spin-up to $P_{\rm eq}$~ is 
\begin{equation}
\Delta (M/\ms) \approx 0.1(P_{\rm eq}/1.5\, {\rm ms})^{-\frac{4}{3}}.
\end{equation} 
 
For $B_{11} = 1, 10$ and $\md_{11} = 1000$, the equilibrium periods are
0.16 and 1.17 s, respectively. The amount of matter required for accretion
is for these two cases $\Delta M = 1.9 \times 10^{-4}$ \ms\ and $1.4
\times 10^{-5}$ \ms. These amounts may be easily accreted during the
common envelope stage.  After the common envelope is dispersed, the NS
appears to be immersed in the strong wind of WR star. Its initial
rotational period is determined by Eq. (6) for accretion at $\md_{\rm
Edd}$. Following Illarionov \& Sunyaev (1976)~ (see also Lipunov 1982),
accretion onto the surface of a rotating neutron star is possible only if
its rotational period is $P_{\rm rot} \apgt P_{\rm eq}^{\prime}$, where
$P_{\rm eq}^{\prime}$ is now an equilibrium period determined by the rate
of accretion from the WR star wind. Then, from Eqs. (2) and (6), for
$\md_{\rm WR} = 4 \times 10^{-8} M_{\rm WR}^{2.5}\, \myr~$ (Langer 1989),
$\md_{\rm Edd} = 1.5 \times 10^{-8}\, \myr$, it follows, that accretion is
possible only for systems with \begin{equation} P_{\rm orb} \aplt 7.3
B_{11}^{-\frac{3}{2}} M_{\rm NS}^{\frac{11}{4}} M_{\rm WR}^{\frac{15}{8}}
M_{\rm tot}^{-\frac{1}{2}} P_{\rm rot}^{\frac{7}{4}} v_{1000}^{-3}~~{\rm
hr} \end{equation} or \begin{equation} P_{\rm orb} \aplt 0.46 M_{\rm
NS}^{\frac{3}{2}} M_{\rm tot}^{-\frac{1}{2}} M_{\rm WR}^{\frac{15}{8}}
v_{1000}^{-3}~~{\rm hr}, \end{equation} where $M_{\rm tot} = M_{\rm NS} +
M_{\rm WR}$. For a higher $P_{\rm orb}$ a neutron star would act as a
``propeller''.  Note, that the critical value of $P_{\rm orb}$ does not
depend on $B$ [in eq.(9)] after elimination of the equilibrium period of
rotation in the common envelope. For $v_{1000} = 1~ {\rm and}~ 1.5$~ the
limiting $P_{\rm orb}$~ is plotted in Fig. 2 [panels (a) and (b)]. It is
evident that for the overwhelming majority of NS + WR systems the
``propeller'' effect may prevent accretion, hence, X-rays emission.  This
result does not depend on the value of \ace\ assumed in the population
synthesis calculations, because the lower limit of the orbital periods of
systems under consideration is determined by the requirement of
accomodation of the WR star in the post-common-envelope orbit. Application
of other prescriptions for mass and momentum loss in the evolution of
close binaries cannot change this result either, because the only
difference which can be expected is a different distribution over orbital
periods with the same minimum period [see Portegies Zwart \& Verbunt
(1995) for a discussion of the influence of different population synthesis
``recipes'']. 

A neutron star in the ``propeller'' stage would experience 
spin deceleration. The characteristic time of a neutron star spin-down 
is (Illarionov \& Sunyaev 1975)
\begin{equation}
t_{\rm A} \approx 4 \times 10^8 B_{\rm 11}^{-\frac{3}{7}} \md_{\rm 
11}^{-\frac{11}{14}} M_{\rm 
NS}^{-\frac{1}{7}} v_{\rm 1000}^{\frac{1}{2}}~~~{\rm yr}
\end{equation} 
or
\begin{equation}
t_{\rm A} \approx 3.8 \times 10^6 B_{\rm 11}^{-\frac{3}{7}} 
M_{\rm NS}^{-\frac{12}{7}} M_{\rm 
WR}^{-\frac{55}{28}} M_{\rm tot}^{-\frac{11}{49}} P_{\rm 
orb}^{\frac{22}{21}} v_{\rm 1000}^{\frac{45}{14}}~~~{\rm yr}.
\end{equation} 
This deceleration time can be compared with a helium  burning time 
in the core of a WR star for which the fit to 
the  \P (1971) and Iben \& Tutukov (1985) data gives:
\begin{equation}
\lefteqn{\log t_{\rm He} \approx 7.15 - 3.7 y + 2.23 y^{1.37}~~{\rm yr},}
\end{equation}
where
$y = \log (M_{\rm WR} / \ms).$~ 
The spin of the neutron star decelerates enough to allow accretion if
orbital period of the system is
\begin{equation}
P_{orb} \aplt 0.5 t_{\rm He}^{\frac{21}{22}} B_{\rm 11}^{\frac{9}{22}} 
 M_{\rm WR}^{\frac{15}{8}} M_{\rm 
tot}^{\frac {3}{14}} v_{\rm 1000}^{-\frac{135}{44}}~~{\rm hr},
\end{equation}
where $t_{He}$~ is in units of $10^6$~ yr. This critical period is 
plotted in Fig. 2 [panels 
(c) and (d)] for WR wind velocity 1500 \kms\ and three values of the 
strength of the magnetic field which fit the range expected   
for neutron stars of the age of several Myr, typical of NS + WR 
systems. Figure 2 shows that even if deceleration is taken into account, 
the 
propeller effect excludes accretion for the overwhelming majority of  
NS +WR  binaries. 

If a WR star forms due to wind mass loss, one can expect a much lower
accretion rate onto NS than in the common envelope, hence, a longer
equilibrium period. For the same wind velocities, conditions for disk
formation may become more favourable.  However, as Fig.~2
clearly shows, descendants of very massive stars probably comprise a
small fraction  of all WR stars in NS + WR binaries: \eg\
$\log M_{WR}/\ms = 1.2$ under our formalism corresponds to a 
main-sequence mass of $\sim 35 \ms.$~

\subsection{Future evolution of the Cyg X-3 system}

Accepting that Cyg X-3 contains a 10 \ms\ WR star and a 10 \ms\ black hole
it is possible to speculate about possible endpoints of its evolution. A
substantial amount of mass may be lost in the WR evolutionary stage. This
matter leaves the system in an isotropic, high-velocity wind, carrying
away specific angular momentum of the primary. Then 
\begin{equation}
\frac{\dot{a}}{a}=-\frac{\dot{M}_{\rm WR}}{\mbh + M_{\rm WR}}.
\end{equation} 
For a 10 \ms\ helium star, the evolutionary
timescale is $\sim 5.5 \times 10^5$ yr (Langer 1989). At the end of 
the evolution of the present WR star,
system will have $P_{\rm f} \approx 8$ hr,
$M_{\rm WR,f} \approx  5 \ms$. If the outcome of evolution is  
a collapse into a  black hole, 
then the system's further evolution  will proceed due to gravitational wave
radiation and in $\sim 4 \times 10^8$ yrs time the system will merge. If a
Wolf-Rayet star forms  a neutron star and expels its
envelope, then, depending on  the magnitude and 
direction of the natal kick, its semi-major axis
may become so large that the neutron
star will not merge with its black hole companion in  a Hubble time. 

\section{Discussion and conclusion}

Our study, as one could expect  from the previous studies of 
a similar kind, suggests that there may 
exist a significant ($\sim 100$ objects) Galactic population of massive 
($\sim 10 \ms$) black 
holes and neutron stars accompanied by Wolf-Rayet stars. Yet, the only 
 example of such systems observed to-date is Cyg X-3 which is suspected to 
harbour a black hole. Rather rough estimates made in Sec. 2~ 
suggest a combination of reasons for the low proportion of such observed 
systems. First, both BH+WR and NS+WR systems 
predominantly form with their orbital periods exceeding several days. 
Second, for disk accretion from  WR star wind to occur, the angular 
momentum of the wind matter has to be high. Hence, the formation of 
accretion 
disks is possible only in BH+WR systems with relatively short orbital 
periods. However, the number of such systems is vanishingly small (Fig. 1). 
Cyg X-3 may be a unique example of such a system.

The formation of accretion disks in NS+WR systems is excluded by their
specific evolutionary history. ~Namely, the spin-up of neutron stars due
to Eddington rate accretion in common envelopes which accompany the birth
of the overwhelming majority of WR stars suffices to prevent accretion via
the action of the ``propeller'' mechanism, if in the subsequent stage, the
wind velocity of the WR star exceeds $\sim 1000\, \kms$~ (Fig. 2), the
actual lower limit for WR star winds.  The lower limit of the model
orbital periods of BH/NS+WR systems depends solely on the radii accepted
for helium stars and does not depend on other parameters of population
synthesis models, such as the common envelope parameter. 
                            
Recently, 
the possibility that Cyg X-3 may harbour a black hole as the compact
object has been discussed by Brown(1995).
Having considered the masses of components in this system, Brown 
(1995) suggests that the progenitor of the present compact object had 
$M_{\rm ZAMS} \apgt 35$ \ms. The immediate progenitor of Cyg X-3 in Brown's 
model contains two massive WR stars.  However, the requirement of 
accomodation of two WR stars in the 
post-common-envelope orbit combined with severe mass 
loss by helium stars prevents, in this model, formation of BH+WR systems 
with orbital periods less than several days.

It is also suggested that low-mass black holes in close binaries may
originate as a result of hypercritical accretion onto neutron stars in
common envelopes (Zel'dovich \etal 1972; Chevalier 1993;  Brown 1995;
Fryer \etal 1996). However, it is unclear whether these black holes may
survive in the common envelopes and avoid swallowing of whole envelope of
the companion. 

All low-mass X-ray binaries with suspected black hole components (with a
possible exception of {\mbox GX 339-4}, Makishima \etal 1986) are soft
X-ray transients (SXT), while SXT are absent among massive X-ray binaries.
Since Cyg X-3 has a short orbital period, similar to the periods of SXT's,
one may wonder why Cyg X-3 is a persistent source. Recently, van Paradijs
(1996) and King \etal (1996, 1997) have discussed the conditions necessary
for disk instability in low-mass X-ray binaries in the presence of X-ray
irradiation. King \etal (1997) have realized that irradiation is much
weaker if the accreting object is a black hole rather than a neutron star,
since black holes do not have hard surfaces and cannot act as
point-like central sources. According to King \etal, a system is  
persistent in X-rays if 
\begin{equation}
\md \apgt \md_{\rm crit}^{\rm irr} \approx 2.86 \times 10^{-11} 
\mbh^{\frac{5}{6}} M_2^{-\frac{1}{6}} P_{\rm hr}^{\frac{4}{3}}~~\myr,~
\end{equation}      
where $M_2$ is the mass of the companion to  the BH.  For $\mbh = M_2 = 10 
\ms$, 
$P_{\rm hr} = 4.8$,  $\md_{\rm crit}^{\rm irr} \approx 10^{-9} \myr$, 
well below the estimated accretion rate in Cyg X-3 of $\md_{\rm acc} 
\approx 5.7 \times 10^{-7} \myr$.
  
\begin{acknowledgements}
The authors acknowledge  support to E.E. through the ``Fund to advance 
the participation of women in higher academic positions'' and to L.R.Y. 
through NWO Spinoza Grant to E. P. J. van den Heuvel.
The present work was completed thanks to the Estonian Science Foundation grant 2446 and 
to the Russian Foundation for Basic Research grant 96-02-16351.  

Both authors acknowledge the warm hospitality extended to them at the
Astronomical Institute ``Anton Pannekoek'' where this study was
accomplished. We are indebted to Prof. E.~P.~J.~van den Heuvel and Dr.~
S.~F.~Portegies~Zwart for their stimulating discussions on the subject
 of the evolution of binaries, and to
the referee, Dr. A. Moffat, for his comments on the earlier version
of the paper.    
\end{acknowledgements}

\end{document}